\documentclass[onecollarge,final]{svjour2}[]
\usepackage{amssymb}
\usepackage{mathrsfs}
\begin{document}
\title{de Bruijn-type identity for systems with flux}
\author{Takuya Yamano} 
\institute{Takuya Yamano
\at Department of Mathematics and Physics, Faculty of Science, 
Kanagawa University, 2946, 6-233 Tsuchiya, Hiratsuka, Kanagawa 259-1293, Japan\\
\email{yamano@amy.hi-ho.ne.jp}
}
\maketitle

\begin{abstract}
We show that an information-theoretic relation called the de Bruijn-type identity 
can be reformulated in a physical context with probability currents. The time derivatives 
of relative entropies under the continuity equation are presented, which shows that 
the conservation of distance between a pair of distributions is generally not guaranteed. 
As an important implication of these results, we discuss and present a possible conceptual 
framework for the classical no-cloning (deleting) theorem and qualitatively assert that 
we can attribute the perfect performance of the operating machine to the openness 
(non-vanishing flow at boundaries between the processing machine and the system) during 
the process. 
\end{abstract}

\section{Introduction}
Since the advent of information theory, it has played a profound role in statistical 
physics. The investigation of proven relations in information theory could facilitate 
further understanding of the laws of physics. In this study, we focus on the one of 
the salient identities that is widely cited in connection with entropy of random 
variables.\\

When an arbitrary random variable $X$ with finite variance is perturbed by another 
independent standard Gaussian random variable $Z$ with zero mean and unit variance but 
scaled by $\sqrt{t}$, the derivative of the associated entropy $H(X+\sqrt{t}Z)$ 
with respect to $t$ has an information relation:
\begin{eqnarray}
\frac{d}{dt}H(X+\sqrt{t}Z)=\frac{1}{2}J(X+\sqrt{t}Z),
\end{eqnarray}
where $J(X)$ denotes the Fisher information associated with the probability distribution 
$f(x)$ for $X$, which is defined as $J(X)=\int f|\nabla \ln f|^2 dx$. The entropy is of 
the Shannon's form $H(X)=-\int f\ln fdx$. This relation is called the de Bruijn identity 
\cite{Stam,Cover,Dembo}. The Fisher information $J(X)$ is obtained in the limiting case 
$t\to 0+$ of the left-hand side of Eq. (1). The factor $1/2$ is included by convention, 
and we will see later that it is absorbed into the diffusion coefficient so that it is 
not essential in our subsequent consideration. In the limiting case $t\to 0+$, this 
identity is generalized to the non-Gaussian variable for $Z$ in Ref. \cite{Naray}. \\

Recently, Guo \cite{Guo} extended this identity to the relative version, 
which is called the de Bruijn-type identity:
\begin{eqnarray}
\lim_{t\to 0+}\frac{d}{dt}KL(\hat{T}_t f\|\hat{T}_tg)=-\frac{1}{2}I(f\|g)
\end{eqnarray}
where $\hat{T}_t$ denotes the perturbation operator acting on the distribution. As a 
special case, $\hat{T}_t$ represents the heat semigroup. The quantities $KL(f\|g)$ and 
$I(f\|g)$ are the Kullback--Leibler distance \cite{KL} (or relative entropy) and the 
relative Fisher information \cite{Blower,Villani} between two distributions $f$ and $g$, 
respectively. When $\hat{T}_t$ denotes a heat semigroup, the distribution maintains the 
Gaussian form over time (Gaussian perturbation). We will not necessarily impose this 
restriction of the Gaussian perturbation for $\hat{T}_t$ in the present paper.

In another recent work, in the context of the minimum mean-square error in estimation 
theory, Verd{\'u} showed that two equal-time Gaussian distributions satisfy the 
following identity \cite{Verdu}:
\begin{eqnarray}
\frac{d}{dt}KL(\hat{T}_t f\|\hat{T}_tg)=-\frac{1}{2}I(\hat{T}_t f\|\hat{T}_tg).
\label{eqn:dBt}
\end{eqnarray}
Also, Hirata {et al.} provided an elementary proof for this identity by way of integration 
by parts \cite{Yoshida}. As an immediate consequence, an integral representation of the KL 
distance in terms of the relative Fisher information follows 
$KL(f\|g)=1/2\int_0^\infty I(\hat{T}_t f\|\hat{T}_tg) dt$ by using the indistinguishability 
property, i.e., $KL(\hat{T}_t f\|\hat{T}_tg)\to 0$ as $t\to\infty$ \cite{Verdu,Yoshida}. 
These above-described intimate relations between relative entropy and the relative Fisher 
information have not been fully investigated in a physics context. 
Since the original de Bruijn identity involves random variables, the dynamics behind 
the probability distribution is by necessity the heat equation. However, there are  
many probability density functions that do not generally follow the heat equation. To 
deal with such cases (i.e., extensions of a de Bruijn-type identity) is our motivation. 
We should also note that the time derivative of a wider class of divergence measures 
(relative entropies) exactly vanishes for probability distributions that follow the 
Liouville equation. This fact has been documented in Ref. \cite{Mackey}.\\
 
In this paper, therefore, we consider the time change of relative entropies between two 
probability distributions, both of which follow the same evolution dynamics. In the de 
Bruijn-type identity, the parameter $t$ denotes the variance of the distribution of 
$\sqrt{t}Z$. In a diffusion process described by the heat equation, the variance of the 
probability distribution is proportional to time. Therefore, the derivative with 
respect to the variance refers to the time derivative. Hence, our first main aim is 
to see how the time derivative of the distance measure between two perturbed 
distributions can be expressed in terms of those distributions. To be precise, it is 
shown that when a system has a probability current, we have a corresponding expression 
of the de Bruijn-type identity. We use the term {\it de Bruijn-type identity} in the 
sense of Eq.(\ref{eqn:dBt}), i.e., the time derivative without taking limit $t\to 0+$. 
We note that readers should distinguish clearly between the original de Bruijn and 
de Bruijn-type identities, since our focus is to present an extended form of the latter 
for nonequilibrium dynamics represented by the continuity equation. 
The continuity equation incorporates the probability current of the system so that 
our formulation can provide the expressions in a general setting, i.e., the de Bruijn-type 
identity for systems with flux. This formulation can also provide a physical 
interpretation to the de Bruijn-type identity that holds for a system, other than a 
heat equation. These investigations strongly encourage us to further explore the links 
between information-theoretic quantities and nonequilibrium physical laws. \\

We also discuss the potential implication of the counterintuitive and 
apparently contradictory conclusion that in the literature about 
the existence of a classical counterpart to quantum-impossible processes 
(i.e., the no-cloning theorem) \cite{Daff,Plastino,Walker,TY4}. Its validity has been 
based on the assumption that both distributions to be compared and the associated 
flux vanish at the boundaries of the system. This premise might not be true in the 
copying process when the physical resources are allowed to interact with the machine. 

The organization of the present paper is as follows. We consider the time changes in  
relative entropies under several dynamics and introduce the relative Fisher information 
in Sec.2. In this section, we also present the de Bruijn-type identities for systems with 
flux in terms of probability current. In Sec.3, we discuss the possible implications 
of the results to the classical analogue of quantum-impossible processes. The final 
section provides the concluding remarks. 
\section{Change of relative entropies in time}
First, we briefly review the form of the relative Fisher information. 
It appears in the de Bruijn-type identity and can be defined 
for two probability distributions $\mathcal{P}_1$ and $\mathcal{P}_2$ as
\begin{eqnarray}
I(\mathcal{P}_1\|\mathcal{P}_2)=\int \mathcal{P}_1\Big| \nabla \ln\frac{\mathcal{P}_1}
{\mathcal{P}_2}\Big|^2dx.\label{eqn:relF}
\end{eqnarray}
This form of definition can be seen in Refs. \cite{Blower,Villani}, but to the best of 
our knowledge, its physical meaning and implication have not been fully explored. The 
identity Eq.(\ref{eqn:dBt}) indicates that the shrinking rate of the distance between two 
distribution functions in terms of the KL distance is equivalent to this information. 
Recall that if we replace $\mathcal{P}$ within the logarithm in the negative Shannon 
entropy with $\mathcal{P}_1/\mathcal{P}_2$, we obtain the KL distance from $\mathcal{P}_1$ 
to $\mathcal{P}_2$, i.e., $KL(\mathcal{P}_1\|\mathcal{P}_2)$. We can obtain the relative 
Fisher information in a similar manner, i.e., if we replace $\mathcal{P}$ within the 
logarithm in the Fisher information $\int \mathcal{P}|\nabla\ln \mathcal{P}|^2dx$ with 
$\mathcal{P}_1/\mathcal{P}_2$, we obtain Eq.(\ref{eqn:relF}). In this sense, the definition 
of the form seems natural. The non-negativity of $I(\mathcal{P}_1\|\mathcal{P}_2)$ is evident 
from the definition, and it can reach zero if and only if the two distributions are identical. 
The quantity $\nabla \ln \mathcal{P}$ is referred to as the score function. Therefore, 
$I(\mathcal{P}_1\|\mathcal{P}_2)$ is regarded as the average of the squared difference of 
two score functions. In application, this information measure is used to 
understand the behaviors of the atomic density profiles \cite{relF3}.\\

Note that so far and hereafter, we omit the argument of the coordinate and time variables as 
$\mathcal{P}=\mathcal{P}(\vec{x},t)$ for simplicity, and we assume that $\mathcal{P}_1$ is 
absolutely continuous with respect to $\mathcal{P}_2$ for $t>0$. Furthermore, the relative 
probability $\mathcal{P}_1/\mathcal{P}_2$ is expected to be bounded on ${\mathbb R}^n$. 
To deal with general nonequilibrium processes, we adopt the following continuity equation: 
\begin{eqnarray}
\frac{\partial \mathcal{P}}{\partial t}=-\nabla\cdot \vec{j}.\label{eqn:cont}
\end{eqnarray}
We now focus on the time derivative of relative distances under 
Eq.(\ref{eqn:cont}) and its particular instances.
\begin{itemize}
\item KL distance $KL(\mathcal{P}_1\|\mathcal{P}_2)$\\
The time derivative is given as 
\begin{eqnarray}
\frac{d}{dt}KL(\mathcal{P}_1\|\mathcal{P}_2)=\int d\vec{x} (\partial_t\mathcal{P}_1)
\ln \frac{\mathcal{P}_1}{\mathcal{P}_2}+\int d\vec{x} \mathcal{P}_2
\left( \frac{(\partial_t\mathcal{P}_1)}{\mathcal{P}_2}-\frac{\mathcal{P}_1}
{\mathcal{P}_2^2}(\partial_t\mathcal{P}_2)\right).
\end{eqnarray}
Substituting $\partial_t\mathcal{P}_i=-\nabla\cdot \vec{j}_i$ for $i=1,2$ and performing 
integration by parts under the assumption of vanishing surface terms (denoted by the 
symbol $s$),
\begin{eqnarray}
\vec{j}_1\ln\frac{\mathcal{P}_1}{\mathcal{P}_2} \Big|_s=0, \quad  
\vec{j}_2\frac{\mathcal{P}_1}{\mathcal{P}_2} \Big|_s=0,\label{eqn:bc}
\end{eqnarray}
we have the expression
\begin{eqnarray}
\frac{d}{dt}KL(\mathcal{P}_1\|\mathcal{P}_2)=\int \mathcal{P}_1\nabla\left( \ln\frac{\mathcal{P}_1}
{\mathcal{P}_2}\right)\left( \frac{\vec{j}_1}{\mathcal{P}_1}-\frac{\vec{j}_2}
{\mathcal{P}_2}\right)d\vec{x}.\label{eqn:dKLdt}
\end{eqnarray}
If we choose the probability current as $\vec{j}_i=\vec{v}\mathcal{P}_i$ for $i=1,2$ with 
common velocity $\vec{v}=d\vec{x}/dt$, which corresponds to the Liouville dynamics, then the time 
change of the distance measure exactly vanishes. This fact is consistent with the property of 
conservation of the KL distance under the Liouville dynamics demonstrated clearly in Ref. \cite{Daff}. 

We note that the zero flux $\vec{j}_i=0$ also provides the constancy of the distance. However,  
we exclude such a stationary circumstance in our consideration. In a particular case, 
systems that follow from the Fick's law $\vec{j}_i=-\mathscr{D}\nabla \mathcal{P}_i$ ($i=1,2$) 
with the diffusion constant $\mathscr{D}$ (i.e., the heat equation) recover the usual 
de Bruijn-type identity:
\begin{eqnarray}
\frac{d}{dt}KL(\mathcal{P}_1\|\mathcal{P}_2)&=&-\mathscr{D}\int\mathcal{P}_1\Big| 
\nabla\left( \ln\frac{\mathcal{P}_1}{\mathcal{P}_2}\right)\Big|^2d\vec{x}\nonumber\\
&=& -\mathscr{D}I(\mathcal{P}_1\|\mathcal{P}_2).\label{eqn:usual}
\end{eqnarray}
This is also true for systems governed by the linear Fokker--Planck equation with a potential 
$\phi(x)$ and the corresponding probability currents:
\begin{eqnarray}
\vec{j}_i = -\frac{d\phi(x)}{dx}\mathcal{P}_i-\mathscr{D}
\frac{\partial \mathcal{P}_i}{\partial x},\quad (i=1,2).
\end{eqnarray}
Since the factor of the difference appearing in Eq.(\ref{eqn:dKLdt}) reads as: 
\begin{eqnarray}
\frac{\vec{j}_1}{\mathcal{P}_1}-\frac{\vec{j}_1}{\mathcal{P}_1}=-\mathscr{D}
\frac{\partial}{\partial x}\left(\ln\frac{\mathcal{P}_1}{\mathcal{P}_2}\right),
\end{eqnarray}
the de Bruijn-type identity immediately follows. 
Note that the appended factor $1/2$ in the original expression Eq.(\ref{eqn:dBt}) is replaced by 
the diffusion constant $\mathscr{D}$. In this sense, Eq.(\ref{eqn:dKLdt}) can be regarded as 
a reformulation of the de Bruijn-type identity under the presence of flow. 
\item Overlap distance $D_o(\mathcal{P}_1\|\mathcal{P}_2)$\\
As another class of distances, we consider the overlap, since it generates other Fisher-like 
relative information, as will be explained below. This quantity is referred to as the fidelity 
measure and is also relevant for studying the information-theoretic aspect of general 
probabilistic theories \cite{Zander}. It is defined as 
\begin{eqnarray}
D_o(\mathcal{P}_1\|\mathcal{P}_2):=\int \sqrt{\mathcal{P}_1\mathcal{P}_2}d\vec{x}.
\end{eqnarray}
The time derivative can be written by substituting the continuity equation and performing 
integration by parts under the boundary condition in Eq.(\ref{eqn:bc}),  
\begin{eqnarray}
\frac{d}{dt}D_o(\mathcal{P}_1\|\mathcal{P}_2)&=&\frac{1}{2}\int d\vec{x}\sqrt{\frac{\mathcal{P}_2}
{\mathcal{P}_1}}(\partial_t\mathcal{P}_1)+\frac{1}{2}\int d\vec{x}\sqrt{\frac{\mathcal{P}_1}
{\mathcal{P}_2}}(\partial_t\mathcal{P}_2)\nonumber\\
&=& \frac{1}{2}\int \vec{j}_1 \nabla\left(\sqrt{\frac{\mathcal{P}_2}
{\mathcal{P}_1}}\right)d\vec{x}+\frac{1}{2}\int \vec{j}_2 \nabla\left(\sqrt{\frac{\mathcal{P}_1}
{\mathcal{P}_2}}\right)d\vec{x}.
\end{eqnarray}
Rearranging the second line, we have the expression  
\begin{eqnarray}
\frac{d}{dt}D_o(\mathcal{P}_1\|\mathcal{P}_2)=-\frac{1}{4}\int d\vec{x} 
\sqrt{\mathcal{P}_1\mathcal{P}_2}\nabla \left( \ln \frac{\mathcal{P}_1}{\mathcal{P}_2}\right)
\left( \frac{\vec{j}_1}{\mathcal{P}_1}-\frac{\vec{j}_2}{\mathcal{P}_2}\right).
\end{eqnarray}
Similarly, in the case of the Liouvillian dynamics, we have exactly the vanishing time derivative. 
However, for a system with the heat equations $\vec{j}_i=-\mathscr{D}\nabla \mathcal{P}_i$ 
($i=1,2$), we have  
\begin{eqnarray}
\frac{d}{dt}D_{o}(\mathcal{P}_1\|\mathcal{P}_2)=-\frac{\mathscr{D}}{4}\int 
\sqrt{\mathcal{P}_1\mathcal{P}_2}\Big|\nabla
\left( \ln\frac{\mathcal{P}_1}{\mathcal{P}_2}\right)\Big|^2d\vec{x}.\label{eqn:odBt}
\end{eqnarray}
This identity is a counterpart of the de Bruijn-type identity. The right-hand side is not the same as 
$I(\mathcal{P}_1\|\mathcal{P}_2)$, but we refer it to a variant of the relative Fisher information 
here, denoting 
\begin{eqnarray}
I_m(\mathcal{P}_1\|\mathcal{P}_2)=\frac{1}{4}\Big\langle \sqrt{\frac{\mathcal{P}_2}
{\mathcal{P}_1}}\Big| \nabla \ln\frac{\mathcal{P}_1}{\mathcal{P}_2}\Big|^2\Big\rangle_{\mathcal{P}_1}.
\end{eqnarray}
It is symmetric $I_m(\mathcal{P}_1\|\mathcal{P}_2)=I_m(\mathcal{P}_2\|\mathcal{P}_1)$ under the 
change of the argument, indicating its pertinence as a potential utility for a distance measure. 
\item A general distance $G(\mathcal{P}_1\|\mathcal{P}_2)$\\
As a demonstration in the general setting, we consider a generic distance measure that can be 
specified by a convex function (a.k.a. the Csisz{\'a}r--Morimoto divergence \cite{Morimoto,Csi}), 
\begin{eqnarray}
G(\mathcal{P}_1\|\mathcal{P}_2):=\int \mathcal{P}_1 \chi
\left( \frac{\mathcal{P}_1}{\mathcal{P}_2}\right)d\vec{x},
\end{eqnarray}
where $\chi$ as a function of the relative density can be taken arbitrarily if  
the property $\chi(1)=0$ is satisfied. Then, the time derivative becomes 
\begin{eqnarray}
\frac{d}{dt}G(\mathcal{P}_1\|\mathcal{P}_2)&=&\!\!\int d\vec{x}(\partial_t\mathcal{P}_1)
\chi\left( \frac{\mathcal{P}_1}{\mathcal{P}_2}\right)+\int d\vec{x}\mathcal{P}_1\chi^\prime \partial_t
\left( \frac{\mathcal{P}_1}{\mathcal{P}_2}\right)\nonumber\\
&=& \!\!\int \vec{j}_1\chi^\prime \nabla\left(\frac{\mathcal{P}_1}{\mathcal{P}_2}\right)d\vec{x}+\!\!
\int \vec{j}_1 \nabla\left(\frac{\mathcal{P}_1}{\mathcal{P}_2}\chi^\prime\right)d\vec{x}-\!\!
\int \vec{j}_2 \nabla\left[\chi^\prime \left(\frac{\mathcal{P}_1}
{\mathcal{P}_2}\right)^2\right]d\vec{x}.
\end{eqnarray}
Similarly, we have performed integration by parts with vanishing terms at the boundary, as follows:
\begin{eqnarray}
\vec{j}_1\chi\Big|_s=0, \quad \vec{j}_1\frac{\mathcal{P}_1}{\mathcal{P}_2}\chi^\prime \Big|_s=0, \quad  
\vec{j}_2\left(\frac{\mathcal{P}_1}{\mathcal{P}_2}\right)^2\chi^\prime \Big|_s=0.
\end{eqnarray}
After the operations of $\nabla$ in the second and third terms in the last expression of the right-
hand side and by rearranging terms, we have: 
\begin{equation}
\int d\vec{x}\left( \frac{\vec{j}_1}{\mathcal{P}_1}-\frac{\vec{j}_2}{\mathcal{P}_2}\right)\mathcal{P}_1
\left\{2\chi^\prime \nabla \left( \frac{\mathcal{P}_1}{\mathcal{P}_2}\right)+
\frac{\mathcal{P}_1}{\mathcal{P}_2}\nabla \chi^\prime \right\}.
\end{equation}
Thus, we have the following final expression for the derivative: 
\begin{eqnarray}
\frac{d}{dt}G(\mathcal{P}_1\|\mathcal{P}_2)=\int d\vec{x}\mathcal{P}_2\left(\frac{\vec{j}_1}
{\mathcal{P}_1}-\frac{\vec{j}_2}{\mathcal{P}_2}\right)\nabla\left[ \chi^\prime\left( 
\frac{\mathcal{P}_1}{\mathcal{P}_2}\right)^2\right].\label{eqn:dGdt}
\end{eqnarray}
\end{itemize}
We readily find that under the Liouville dynamics $\vec{v}_i=\vec{j}_i/\mathcal{P}_i=\vec{v}$ 
for $(i=1,2)$, a wide class of distance measure specified by the form of the function $\chi$ is 
time invariant, which is consistent with the direct demonstration presented in Ref. \cite{Plastino}. 
In this sense, Eq.(\ref{eqn:dGdt}) provides the de Bruijn-type identity for systems with nonequilibrium 
state, which is widely applicable beyond heat phenomenon. The results obtained above are summarized 
in Table 1. Note that the time derivatives under the continuity equation have a common factor of the 
difference in the flux $\vec{j}_1/\mathcal{P}_1-\vec{j}_2/\mathcal{P}_2$, which is the primary cause 
for the non-vanishing time change.
We note further that the de Bruijn-type identity Eq.(\ref{eqn:usual}) and a variant Eq.(\ref{eqn:odBt}) 
are reminiscent of the H-theorems verified by the Fokker--Planck equations and master equations \cite{Risk}. 
Indeed, since the two relative Fisher information $I$ and $I_m$ are always positive, the distances between 
two probability distributions decrease in time, and it gives the rate at which the distances shrink. 
\begin{table}
 \caption{The de Bruijn type identities for dynamics}
 \begin{center}
  \begin{tabular}{|c||c|c|c|}
    \hline
       &  Liouville eq.  &  Heat eq.  & Continuity eq.\\
    \hline
    $\frac{dD_{KL}}{dt}$   &  0  & $-\mathscr{D}I(\mathcal{P}_1\|\mathcal{P}_2)$  & 
    $\Big\langle \nabla\left( \ln\frac{\mathcal{P}_1}{\mathcal{P}_2}\right)
    \left( \frac{\vec{j}_1}{\mathcal{P}_1}-\frac{\vec{j}_2}{\mathcal{P}_2}\right)
    \Big\rangle_{\mathcal{P}_1}$ \\
    \hline
    $\frac{dD_{O}}{dt}$   &  0  & $-\mathscr{D}I_m(\mathcal{P}_1\|\mathcal{P}_2)$ &  
    $\Big\langle \frac{\sqrt{\mathcal{P}_1\mathcal{P}_2}}{4} \nabla
    \left( \ln \frac{\mathcal{P}_1}{\mathcal{P}_2}\right)
    \left( \frac{\vec{j}_1}{\mathcal{P}_1}-\frac{\vec{j}_2}{\mathcal{P}_2}\right)
    \Big\rangle_{\mathcal{P}_1}$ \\
    \hline
    $\frac{dG}{dt}$   &  0  & $-\mathscr{D}\Big\langle \nabla
    \left( \ln \frac{\mathcal{P}_1}{\mathcal{P}_2}\right)
    \nabla \left[ \chi\prime\left( \frac{\mathcal{P}_1}{\mathcal{P}_2}\right)^2\right]
    \Big\rangle_{\mathcal{P}_2}$  & 
    $\Big\langle \nabla \left[ \chi\prime\left( \frac{\mathcal{P}_1}
    {\mathcal{P}_2}\right)^2\right] \left( \frac{\vec{j}_1}{\mathcal{P}_1}-\frac{\vec{j}_2}
    {\mathcal{P}_2}\right)\Big\rangle_{\mathcal{P}_2}$ \\
    \hline
  \end{tabular}
 \end{center}
\end{table}
\section{On the relevance to the classical no-go theorem}
The derived de Bruijn-type identities derived indicate that distance is not generally 
conserved under flow dynamics. We discuss here a significant implication of this 
fact to the classical version of the fundamental property of quantum mechanics, i.e., the 
impossibility of copying (or deleting) a source state into (or from) a target system 
\cite{No} without perturbing it. We refer to both operations as broadcasting, but for the 
sake of simplicity, we hereafter only consider copying (cloning). 
The no-cloning theorem has traditionally been considered a genuine manifestation of the 
non-classical effect that appears in quantum information processing. It forbids us to make an 
identical copy of the original quantum state by a unitary operation without destroying the 
original \cite{No} (for later advances, see e.g. \cite{RMP}). In Refs. \cite{Daff,Plastino}, 
it is pointed out that cases of continuous variables under certain conditions also follow the 
same theorem. The counterpart to this quantum feature in classical processing, 
if it exists, is an amazing and counterintuitive idea. Therefore, a compromise (coarse-grained) 
method to this assertion was presented in Ref. \cite{Walker}, where input distributions with 
nonzero resolution (see Appendix) under Liouville dynamics may be broadcast with fidelity 
arbitrarily close to unity \footnote{In this sense, the authors of Ref. \cite{Daff,Plastino} 
only refer to infinite resolution broadcasting or perfect accuracy.}. 
These considerations are significant for exploring the border between classical and quantum 
physics, and it would be particularly reasonable to address this issue here in view of its 
high relevance to our derived de Bruijn-type identities. This resolution-based argument is, 
however, subtle, and we must be circumspect about it in the following sense: 
\begin{itemize}
\item 
The machine's resolution $\epsilon_{machine}$ is fixed before the operation and is supposed 
to be smaller than the one determined by the input, i.e., $\epsilon [p] > \epsilon_{machine}$ 
is assumed. However, when the converse holds for some input distributions, we do not achieve 
high fidelity, as claimed in Ref. \cite{Walker}. 
\item 
We normally do not know and cannot know the initial source state {\it a priori} without 
observing it. Accordingly, we cannot prepare the resolution of the copying machine 
corresponding to the inputs prior to the operation. The resolution must be set before 
observation (and subsequent copying) and not the other way around. 
\end{itemize}
Without getting involved in the above incommensurate aspects of resolution, we will show that 
it is possible to interpret the issue. We think that the non-conservation of distance measures 
demonstrated in the classical copying processes \cite{Daff,Plastino,TY4} has the following 
interpretation. Recall that our argument presented so far has relied heavily on the 
premise of vanishing surface terms, i.e., probabilities and flow vectors disappearing at the 
system's boundary. This assumption makes the integration by parts an effective step towards 
obtaining the time change of distance measures and leads to the physical formulation of the 
de Bruijn-type identities. \\

The gist of the claim of the impossibility performance even in a classical macroscopic 
system lies in the fact that the extra inflow and outflow at the boundary are negligible. 
This allows that the relative entropy employed remains constant before and after the 
transfer process. Therefore, a possible explanation is that the source and the target 
systems become temporarily open systems mediated by the machine in the course of the 
process. The two non-conservative initial probabilities consequently break the constancy 
of the distance measure, and the probability does not evolve according to the Liouville 
equation. This may suggest a view that the machine works as a probability ''sink'' or 
''bath'' depending on the processing dynamics. A similar qualitative explanation can 
be found in Ref. \cite{Brody}, where in the derivation of the upper bound for the entropy 
production, the authors mention that an open (quantum) system may entail a flow across 
the boundaries. In general, the machine can have much a larger degree of freedom than 
both the system and the target. Therefore, it works as a heat bath, and the machine's 
source distribution itself will not be perturbed considerably during operation.\\

If we consider the above viewpoint, it makes sense that the time invariance of a wide 
class of relative metrics does not contradict the Liouville evolution of probability 
distributions. Once we consider that the classical cloning process does not follow the 
Liouville dynamics, there may be no need to introduce an artifice such as the resolutions 
of input distributions. 

\section{Conclusion} 
We have derived the time change of the distance measures under general dynamics. 
These identities can be regarded as extensions of the cerebrated de Bruijn-type 
identity, in that the original identity is obtained in a particular case of the heat 
equation. We have emphasized that the invariance of relative entropies in time can 
always be achieved when a system follows the Liouvillian evolution. This fact naturally 
leads to a possible compromise interpretation of the classical impossibility processes 
that is originally inherent in the quantumness of systems. The conservation of  
distance between the input and output states in broadcasting processes is violated except 
for the Liouville dynamics in the sense that the time derivative of the relative entropies 
does not vanish. The existence of a classical version of the no-go theorem can be 
interpreted as non-Liouvillian dynamics of the probability distribution.
As the rate of change of the relative entropy can be interpreted as information loss 
(or gain) across the boundaries in the form of a nonzero flux per unit time, the quantity 
is an indicator of the credibility of classical operations. However, the explanation 
presented is still intuitive, and to make the consideration more qualitative, we will need 
to introduce a refinement of the comparative quantity, say, the partial success rate of 
the broadcast, instead of assuming perfect performance of the machine.  
These investigations will provide a deeper understanding of the functioning of machines 
in dynamical processes and provide a basis for the emergence of classicality from 
quantumness. The classical counterpart of the quantum no-cloning (and no-deleting) 
is not a phantom analogue. 
\section{Appendix} 
A particular type of resolution for probability distributions was introduced to assist 
the success of the near-ideal copying of the classical distributions \cite{Walker}. 
The shape of the distribution to be copied is required to have well-posed features.    
The resolution to distinguish two probability distributions is called the 
$\epsilon$-resolution and is defined as: 
\begin{eqnarray}
\epsilon [p]:= \displaystyle{\max_{\delta}} \frac{2|\delta|}{\int dx|p(x)-p(x-\delta)|}, 
\nonumber
\end{eqnarray}
where $\delta$ is the displacement in position $x$. This resolution has the following  
features:
\begin{itemize}
\item It is essentially equivalent to the inverse of the Lipschitz constant $L_{p(x)}$ 
of the distribution function on a given interval defined as 
\begin{eqnarray}
L_{p(x)}:=\sup_\delta \frac{d(p(x),p(x-\delta))}{\delta},\nonumber
\end{eqnarray}
where the numerator denotes a metric to be measured. The factor $2$ in the case of the 
$\epsilon$-resolution comes from the choice of the metric $d(p,q)$ as the trace norm 
defined as $\int dx|p-q|/2$. Since the Lipschitz continuity requires the distribution 
function to be a stronger (more demanding) condition for smoothness, distributions 
that are not Lipschitz continuous need alternative fine-grained procedures. Just as 
the Lipschitz constant depends on the function, the $\epsilon$-resolution is also 
contingent on the source state.
\item
When the probability distributions are symmetric and monotonically non-increasing 
away from the origin, it is given as the inverse of the height of the distribution: 
$\epsilon[p]=1/p(0)$. In many areas of science, where we cannot expect this 
well-posedness in distribution, skewed distributions such as the gamma, Weibull, 
and log-normal distributions, appear. In these cases, we calculate the resolution 
by the definition, since the distributions have zero-height at the origin. 
\item 
When applied to the classical copying scenario, mismatched resolutions i.e., 
$\epsilon [p] < \epsilon_{machine}$ can occur depending on the input distributions, 
since the initial machine's resolution is fixed. These situations could occur when 
we copy different types of input distributions consecutively many times, as we usually 
do in our lives. It is not a trivial matter to iterate Eqs.(26) and (27) of 
Ref. \cite{Walker}. As a consequence, it does not produce as the desired high fidelity. 
The optimal tuning mechanism in the machine side corresponding to each input distribution 
must be introduced.\footnote{In this regard, we mention also that making an operationally 
clear distinction between preparation procedures, transformation, and measurements is 
important for the recent ontic/epistemic discussion of the nature of the quantum state 
(e.g., Ref. \cite{Blasiak} and references therein). It has significance in the study of 
classical analogues of aspects of quantum physics.}
\end{itemize}  
\begin{acknowledgement}
The author wishes to thank H. Yoshida at Ochanomizu University for valuable discussions 
on the de Bruijn identity. The part of this work was presented at the JPS (Japan 
Physical Society) annual meeting held at the Hiroshima University, 26 Mar. 2013.
\end{acknowledgement}
 

\begin{thebibliography}{99}
\bibitem{Stam} 
A.J. Stam, Inf. Contr. {\bf 2} 101 (1959).

\bibitem{Cover} 
T. Cover \and J. Thomas, {\it Elements of Information Theory 2nd ed.}
Wiley-Interscience (2006). 

\bibitem{Dembo} 
A. Dembo \and T. Cover \and J. Thomas, 
IEEE Trans. Inform. Theory, {\bf 37} 1501 (1991).

\bibitem{Naray} 
K.R. Narayanan \and A.R. Srinivasa, 
"On the thermodynamic temperature of a general distribution", eprint arXiv/0711.1460v2.

\bibitem{Guo} 
D. Guo, Proc. IEEE Int. Symp. Inf. Theory Seoul Korea, Jun. 28-Jul. 3,  814 (2009).

\bibitem{KL} 
S. Kullback \and R.A. Leibler, 
Ann. Math. Stat. {\bf 22} 79 (1951); 
S. Kullback, 
{\it Information Theory and Statistics}, Wiley, New York (1959).

\bibitem{Blower} 
G. Blower, 
{\it Random Matrices: High Dimensional Phenomena, ser. London Mathematical Society Lecture Notes.} 
Cambridge Univ. Press. Ch.6, p.211, (2009).

\bibitem{Villani} 
C. Villani, {\it Topics in Optimal Transportation, Graduate Studies in Mathematics} 
Vol.58, American Mathematical Society, p.278 (2000). 

\bibitem{Verdu} 
S. Verd{\'u}, 
IEEE Trans. Inform. Theor. {\bf 56} 3712 (2010). 

\bibitem{Yoshida} 
M. Hirata \and A. Nemoto \and H. Yoshida,  
Entropy {\bf 14} 1469 (2012).

\bibitem{Mackey} 
M.C. Mackey, Rev. Mod. Phys. {\bf 61} 981 (1989). 

\bibitem{Daff} 
A. Daffertshofer \and A.R. Plastino \and A. Plastino, 
Phys. Rev. Lett. {\bf 88} 210601 (2002).

\bibitem{Plastino} 
A.R. Plastino \and A. Daffertshofer, 
Phys. Rev. Lett., {\bf 93} 138701 (2004).

\bibitem{Walker} 
T.A. Walker \and S. Braunstein,  
Phys. Rev. Lett. {\bf 98} 080501 (2007).

\bibitem{TY4} 
T. Yamano \and O. Iguchi,  
Europhysics Lett. {\bf 83} 50007 (2008).

\bibitem{relF3} 
J. Antol{\'i}n  \and J.C. Angulo \and S. L{\'o}pez-Rosa, 
J. Chem. Phys. {\bf 130} 074110 (2009); 
P. S{\'a}nchez-Moreno \and A. Zarzo \and J.S. Dehesa,   
J. Phys. A: Math. Theor. {\bf 45} 125305 (2012).

\bibitem{Zander}
C. Zander \and A.R. Plastino, 
Europhysics Lett. {\bf 86} 18004 (2009).

\bibitem{Morimoto} 
T. Morimoto, J. Phys. Soc. Jpn. {\bf 12} 328 (1963). 

\bibitem{Csi} 
I. Csisz{\'a}r, Stud. Math. Hung. {\bf 2} 299 (1967). 

\bibitem{Risk}
H. Risken, {\it The Fokker--Planck Equation}, Springer, New York, (1989).

\bibitem{No} 
W.K. Wootters \and W.H. Zurek, 
Nature, {\bf 299} 802 (1982); 
D. Dieks, Phys. Lett. A {\bf 92} 271 (1982).

\bibitem{RMP} 
V. Scarani \and S. Iblisdir \and N. Gisin \and A. Ac{\' i}n,  
Rev. Mod. Phys. {\bf 77} 1225 (2005).

\bibitem{Brody} 
D. Brody \and B. Meister, 
Phys. Lett. A, {\bf 204} 93 (1995).

\bibitem{Blasiak}
P. Blasiak, 
Phys. Lett. A, {\bf 377} 847 (2013).
\end{thebibliography}
\end{document}